# Transparent FPGA Acceleration with TensorFlow

Simon Pfenning*, Philipp Holzinger* and Marc Reichenbach*
* Department of Computer Science, Chair of Computer Architecture
Friedrich-Alexander-University Erlangen-Nuremberg, Germany
{simon.pfenning, philipp.holzinger, marc.reichenbach}@fau.de

*Abstract*—Today, artificial neural networks are one of the major innovators pushing the progress of machine learning. This has particularly affected the development of neural network accelerating hardware. However, since most of these architectures require specialized toolchains, there is a certain amount of additional effort for developers each time they want to make use of a new deep learning accelerator. Furthermore the flexibility of the device is bound to the architecture itself, as well as to the functionality of the runtime environment.

In this paper we propose a toolflow using TensorFlow as frontend, thus offering developers the opportunity of using a familiar environment. On the backend we use an FPGA, which is addressable via an HSA runtime environment. In this way we are able to hide the complexity of controlling new hardware from the user, while at the same time maintaining a high amount of flexibility. This can be achieved by our HSA toolflow, since the hardware is not statically configured with the structure of the network. Instead, it can be dynamically reconfigured during runtime with the respective kernels executed by the network and simultaneously from other sources e.g. OpenCL/OpenMP.

## I. INTRODUCTION

Modern AI algorithms based on the concept of artificial neural networks (ANN) have significantly expanded the capabilities of machine learning (ML). Recently, there have been increasing efforts to make those achievements accessible to mobile devices as well. However, the power consumption in this environment is severely restricted which makes highly energy-efficient designs inevitable. For this reason, throughput optimized GPGPUs as the past driving force of ML are not able to similarly dominate there in the long term. Instead, application-specific architectures have an increasingly important role already now. Although current developments like Google's TPU are dedicated ASIC solutions, FPGAs are also gaining in significance. Their reconfigurability gives them the flexibility to adapt to a wide range of tasks while retaining a high efficiency. With this capability they can not only accelerate the neural networks themselves but also pre- and post-processing as well as sensor fusion which can highly differ between applications.

Even though previous FPGA approaches have already shown advantages over traditional general purpose architectures, they are not able to fully utilize them in a complete heterogeneous system. They usually claim the hardware exclusively for the ANN or require own frameworks and development flows. This leads to the situation where application developers, who want to use the new hardware capabilities, often need to diverge from their already established workflow.

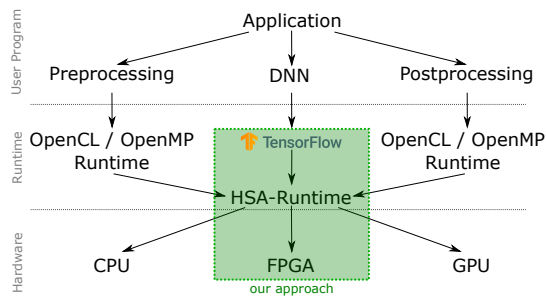

Fig. 1: Mapping of DL applications with our toolchain.

Furthermore, the hardware might not be utilized to its fullest, since the functionality is limited to the processing of certain network types.

Hence, in this paper we introduce a new method for making use of hardware accelerators in the field of deep learning (DL) without requiring a considerable amount of additional effort and yet with a high degree of flexibility. Since TensorFlow (TF) is one of the most commonly used DL frameworks, we decided to use it as the frontend of our toolflow. In order to achieve as much flexibility with the system as possible, the underlying methodology is based on the HSA Foundation standard [1]. This standard is delivering a common way for controlling conforming devices like GPUs, CPUs or DSPs.

## II. RELATED WORK

There are similar approaches which tried to synthesize static netlists out of TF for FPGAs. LeFlow [2] uses the TF internal compiler to obtain a static compute graph of the network, which is then transferred via high level synthesis into an FPGA specific netlist. Xilinx's Vitis AI [3] framework takes a similar way, by analyzing and optimizing the DL model with their own AI Compiler and transferring the result to the Vitis AI Runtime, which is deploying the work to FPGA accelerators. Contrary to these static procedures, our method based on the principles described in [4], [5] does not statically map the model onto the FPGA as a whole. Instead, it utilizes the TF runtime for issuing the workload as pars pro toto. This allows for a more flexible use of the FPGA, not solely for the execution of a specific network model.

## III. CONCEPT

Figure 1 shows the mapping of applications to dedicated hardware with our proposed approach. In contrast to other solutions our concept refrains from using a secondary toolchain which processes the frozen graph for the FPGA. Instead everything needed is completely integrated into TF itself and can be utilized by the same Python/C++ calls developers are already familiar with.

TABLE I: Utilization of the Programmable Logic

| Kernel | LUTs | FFs | BRAM | DSPs |
|---|---|---|---|---|
| Shell | 9915 (14.1%) | 8544 (6.1%) | 10 (4.6%) | 0 (0.0%) |
| Role 1 | 9984 (14.1%) | 8479 (6.0%) | 21 (9.7%) | 22 (6.1%) |
| Role 2 | 9501 (13.5%) | 7851 (5.6%) | 23 (10.6%) | 8 (2.2%) |
| Role 3 | 5091 (7.2%) | 4935 (3.5%) | 21 (9.7%) | 6 (1.7%) |
| Role 4 | 7881 (11.2%) | 7926 (5.6%) | 21 (9.7%) | 12 (3.3%) |

TABLE II: Overhead of FPGA TensorFlow [µs] (n=1000)

| Operation | Occurrence | TensorFlow | HSA Runtime |
|---|---|---|---|
| device/kernel setup | once | 156230 | 39032 |
| reconfiguration | if not configured | 0 | 7424 |
| dispatch latency | every dispatch | 27 | 10 |

TABLE III: Efficiency benefit compared to CPU (n=1000)

| | Role 1 | Role 2 | Role 3 | Role 4 |
|---|---|---|---|---|
| OP/cycle increase | 6.51× | 3.03× | 18.62× | 6.98× |

However, *applications* rarely only contain procedures provided by a DL framework. Instead, they are usually divided into the network inference itself and several external *pre- and post-processing* steps e.g. for data acquisition and sensor fusion. Therefore, it is the responsibility of a comprehensive toolflow to forster these common use cases. Our proposed concept combines these two aspects by abstracting the low level details with a common standard [1] implemented in the *HSA Runtime* and its associated drivers. It manages all HSA devices in the system, informs its users about the status of the underlying hardware and synchronizes dispatched tasks. The necessary HSA runtime calls can be generated either by a standard OpenCL/OpenMP compiler or the TF framework. For this purpose, the TF runtime has been extended by a respective device backend. It detects and manages all the accessible HSA devices visible to the framework. By using an annotation in their Python- or C-Code, developers can induce to execute operations on certain device-types. If TF is able to find a registered kernel implementation for HSA devices it will be dispatched using HSA runtime calls which are made available to TF by our extension.

In principle this is not different to the method for any other accelerator. The major difference here is, in case of an FPGA there are two options of what a registered kernel can be. The simple and most flexible solution would be an OpenCL implementation. During inference this would result in compilation of an intermediate format shared by all HSA devices. After a runtime synthesis the device specific bitstream is generated and deployed to the FPGA.

The great advantage of this method lies in its flexibility, since not only the synthesis target can be changed during runtime, but also the same OpenCL kernel can be used for various types of HSA accelerators. On the downside, this approach leads to a significant increase in runtime and energy costs. Especially due to online synthesis. As we want to focus on a mobile use case, we found it a better solution to register presynthesized bitstreams as kernels for TF, which are then deployed during runtime and used for partial reconfiguration of the FPGA. This way we still maintain a high degree of flexibility without having to suffer from the disadvantage of highly increased energy consumption. Since the FPGA is not configured once with a static network structure, but it is dynamically reconfigured for each kernel call, it is not monopolized by the network and can be used for other tasks like pre- and post-processing steps.

IV. EVALUATION

We ran a preliminary implementation of our concept on an Ultra96 board. Since not all parts are conclusively implemented, the measurements are not to be considered complete, but give a first impression. Table I depicts the FPGA resource utilization of the shell and several layer variants:

1) Fully connected (float32)
2) Fully connected with barrier (float32)
3) Conv 5×5, 1 filter, fixed weights (int16)
4) Conv 3×3, 2 filters, fixed weights (int16)

The overhead caused by our TensorFlow-HSA approach is listed in table II. At the beginning all device management mechanisms are set up for kernel dispatches. This delay therefore occurs only once during execution. Reconfiguration is automatically handled by the runtime and happens every time when a kernel that is not currently loaded on the FPGA is executed. In this process a LRU eviction scheme is used if more roles than available regions need to be handled. TF can consider this trade-off to either generate a lower number of generic roles or fix layer weights to have more efficient hardware. Finally, the kernels can be dispatched with a low latency as often as needed. Our first measurements shown in table III already demonstrate an improvement of up to 18.62× over a plain ARM Cortex A53 implementation. In the future we will further optimize the components to be comparable to state-of-the-art approaches.

ACKNOWLEDGMENT

This work is a result of the project "KI-Flex" (project number 16ES1027), funded by the German Federal Ministry of Education and Research (BMBF) within the founding program Microelectronic from Germany innovation driver.

REFERENCES

[1] HSA Foundation, "HSA Specification Version 1.2," May 2018. [Online]. Available: http://www.hsafoundation.com/standards/
[2] D. Noronha, K. Gibson, B. Salehpour, and S. Wilton, "LeFlow: Automatic Compilation of TensorFlow Machine Learning Applications to FPGAs," 12 2018, pp. 393–396.
[3] I. Xilinx, *Vitis AI User Documentation*, December 2020. [Online]. Available: https://www.xilinx.com/html_docs/vitis_ai/1_3/
[4] M. Reichenbach, P. Holzinger, K. Häublein, T. Lieske, P. Blinzer, and D. Fey, "Heterogeneous Computing Utilizing FPGAs," *J. Sign. Process. Syst.*, vol. 91, pp. 745–757, May 2018.
[5] P. Holzinger, M. Reichenbach, and D. Fey, "A New Generic HLS Approach for Heterogeneous Computing: On the Feasibility of High-level Synthesis in HSA-compatible Systems," in *Proc. 18th Int. Conf. Embed. Comput. Syst.: Arch., Model. and Simul.*, 2018, pp. 18–27.